\documentclass[superscriptaddress,aps,prl,twocolumn,floatfix,noeprint]{revtex4-1}
\usepackage{amsmath}
\usepackage{mathtools}
\usepackage{graphicx}
\usepackage{afterpage}
\usepackage{amssymb}
\usepackage{bbold}

\usepackage{epstopdf}
\usepackage{xcolor}
\usepackage[caption=false,position=top,singlelinecheck=off,justification=raggedright]{subfig}

\usepackage{tikz}

\usetikzlibrary{calc,fadings,decorations.pathreplacing}
\newcommand\pgfmathsinandcos[3]{%
  \pgfmathsetmacro#1{sin(#3)}%
  \pgfmathsetmacro#2{cos(#3)}%
}
\newcommand\LongitudePlane[3][current plane]{%
  \pgfmathsinandcos\sinEl\cosEl{#2} 
  \pgfmathsinandcos\sint\cost{#3} 
  \tikzset{#1/.style={cm={\cost,\sint*\sinEl,0,\cosEl,(0,0)}}}
}
\newcommand\LatitudePlane[3][current plane]{%
  \pgfmathsinandcos\sinEl\cosEl{#2} 
  \pgfmathsinandcos\sint\cost{#3} 
  \pgfmathsetmacro\yshift{\cosEl*\sint}
  \tikzset{#1/.style={cm={\cost,0,0,\cost*\sinEl,(0,\yshift)}}} %
}

\newcommand\DrawLatitudeCircle[2][1]{
  \LatitudePlane{\angEl}{#2}
  \tikzset{current plane/.prefix style={scale=#1}}
  \pgfmathsetmacro\sinVis{sin(#2)/cos(#2)*sin(\angEl)/cos(\angEl)}
  \pgfmathsetmacro\angVis{asin(min(1,max(\sinVis,-1)))}
  \draw[current plane] (\angVis:1) arc (\angVis:-\angVis-180:1);
  \draw[current plane,dashed] (180-\angVis:1) arc (180-\angVis:\angVis:1);
}

\tikzset{%
  >=latex, 
  inner sep=0pt,%
  outer sep=2pt,%
  mark coordinate/.style={inner sep=0pt,outer sep=0pt,minimum size=3pt,
    fill=black,circle}%
}

\newcommand*{\CC}{%
  \textsf{C\kern-1ex C}%
}

\DeclarePairedDelimiter\bra{\langle}{\rvert}  
\DeclarePairedDelimiter\ket{\lvert}{\rangle}      
\DeclarePairedDelimiterX\braket[2]{\langle}{\rangle}{#1 \delimsize\vert #2}          
\begin{document}

\title{Fluctuations and Non-Hermiticity in the Stochastic Approach to Quantum Spins} \author{S. E. Begg} \affiliation{Department of Physics, King's
  College London, Strand, London WC2R 2LS, United Kingdom}
\author{A. G.  Green} \affiliation{London Centre for Nanotechnology,
  University College London, Gordon St., London, WC1H 0AH, United
  Kingdom}\author{M. J. Bhaseen} \affiliation{Department of Physics, King's College London,
  Strand, London WC2R 2LS, United Kingdom} \date{\today}

\begin{abstract}
We investigate the non-equilibrium dynamics of isolated quantum spin
systems \text{via} an exact mapping to classical stochastic differential
equations. We show that one can address significantly larger system
sizes than recently obtained, including two-dimensional systems with
up to 49 spins. We demonstrate that the results for physical
observables are in excellent agreement with exact results and
alternative numerical techniques where available. We further develop a
hybrid stochastic approach involving matrix product states. In the
presence of finite numerical sampling, we show that the non-Hermitian
character of the stochastic representation leads to the growth of the
norm of the time-evolving quantum state and to departures for physical
observables at late times. We demonstrate approaches that correct for
this and discuss the prospects for further development.

\end{abstract}

\maketitle

Experimental progress on cold atomic gases and trapped ions has led to
pristine realizations of isolated quantum spin systems in and out of
equilibrium \cite{Friedenauer2008,Simon2011,Meinert2013,Jurcevic2014}.  
 This has stimulated intense theoretical activity to
expose the unitary dynamics of paradigmatic spin Hamiltonians, with a
view towards extracting universal results \cite{Polkovnikov2011,Eisert2015,Essler2016}. Much of the attention
has focused upon one-dimensional spin models due to the
availability of analytical \cite{Calabrese2012a,Caux2013,Pozsgay2013,Fagotti2014,Piroli2017,Piroli2018} and numerical \cite{Vidal2004,Haegeman} techniques. This has yielded fundamental insights into the nature of
thermalization \cite{Rigol2007,Alba2015,Hallam2019} and to the development of new techniques \cite{Barry2008,Caux2013,Wurtz2018}. The prediction of dynamical quantum phase transitions (DQPTs)
occuring in the time-domain \cite{Heyl} has been confirmed by experiment on
Ising Hamiltonians realized with trapped ions \cite{Jurcevic2017}.  This opens the door
to time-resolved dynamics in tunable quantum spin systems, allowing
direct comparison between theory and experiment.

A recent theoretical approach to non-equilibrium quantum spin systems permits an exact mapping to classical stochastic
differential equations (SDEs)
\cite{Hogan2004,Galitski2011,Ringel2013,DeNicola2019,DeNicola2019long}.
The time-evolution of quantum observables is encoded by classical
averages over independent realizations of the stochastic process.  The
method is therefore inherently parallelizable and can be
implemented by numerically sampling the SDEs
\cite{DeNicola2019,DeNicola2019long}. The stochastic approach is
rather general, since it applies to both integrable and non-integrable
Hamiltonians, including those in higher dimensions. The stochastic
framework also reveals deep connections between classical and quantum
dynamics, as recently illustrated in the context of DQPTs
\cite{DeNicola2019,DeNicola2019long}.

In this work, we show that the stochastic approach to quantum spin
systems can address significantly larger system sizes than 
previously possible \cite{DeNicola2019,DeNicola2019long}. 
This is obtained
through the use of a Heun integration scheme and the
elimination of divergent stochastic trajectories. We show that the
results obtained for the one-dimensional (1D) quantum Ising model are in
very good agreement with those obtained from free fermions \cite{Heyl}
and \textit{via} matrix product operator (MPO) methods \cite{Zaletel2015}. We also provide results for the two-dimensional (2D) quantum Ising model with up to 49 spins.
We relate the growth of stochastic fluctuations at late times to the
non-Hermiticity of the effective stochastic Hamiltonian.  Due to the
impact of finite numerical sampling, this leads to an increase in the
norm of the time-evolving quantum state and to departures for
observables at late times. This can be
  partially corrected by rescaling by the norm.
We show that a hybrid numerical scheme combining SDEs
  with matrix product states can reduce the number of noise variables,
  thereby extending the simulation time.
We conclude and provide directions for research.

{\em Stochastic Approach.}--- Here, we briefly recall the principal steps in the
stochastic approach to quantum spin systems \cite{Hogan2004,Galitski2011,Ringel2013} following the notations
in \cite{Ringel2013,DeNicola2019}. We begin with a generic Heisenberg Hamiltonian
\begin{align} \hat{H} = - \frac{1}{2} \sum_{ijab}  J^{ab}_{ij} \hat{S}^a_i  \hat{S}^b_j - \sum_{ia} h^a_i \hat{S}^a_i ,\label{eq:heisenberg}\end{align}
where $i$ and $j$ indicate lattice sites, $J^{ab}_{ij}$ are exchange interactions, and $h^a_i$ are magnetic fields. 
The spin operators, $\hat{S}^a_i$, obey the canonical commutation relations, $[\hat{S}^a_i,\hat{S}^b_j] = i \epsilon^{abc}\delta_{ij} \hat{S}^c_i$, with $\hbar = 1$ and $a,b,c \in \{x,y,z\}$. 
The corresponding time-evolution operator between times $t_i$ and $t_f$ is of the form $ \hat{U}(t_f,t_i) = \mathbb{T}e^{-i \int_{t_i}^{t_f} \hat{H}(t) dt} $, where $\mathbb{T}$ denotes time-ordering.
The key idea is that the exchange interactions can be decoupled
using a Hubbard--Stratonovich transformation, which introduces fluctuating stochastic fields $\varphi$ \cite{Hogan2004,Galitski2011,Ringel2013}.
The time-evolution operator can be expressed as
\small
\begin{align}  \hat{U}(t_f,t_i)\! =\!  \int \! D\mu(\varphi)  \mathbb{T} \exp \Big[\! -\! i \!\int\limits_{\mathclap{t_i}}^{\mathclap{t_f}}\! dt \sum_{ja} \big(\frac{-1}{\sqrt{i}} \varphi^a_{j} \! - \!h_j^a\big) \hat{S}^a_{j}\Big] , \label{eq:HStransf} \end{align}
\normalsize
with the Gaussian noise measure $D\mu(\varphi) = \prod_{ja}
D\varphi_j^a ~\text{exp}\big(-\frac{1}{2} \int_{t_i}^{t_f} dt
\sum_{ijab} \varphi^a_i~(J^{-1})^{ab}_{ij} \varphi^b_{j}\big) $
\cite{Hogan2004,Ringel2013}.  This formulation describes $N$ decoupled
spins evolving under effective ``magnetic" fields. It is inherently
non-Hermitian due to the factor of $1/\sqrt{i}$, and the presence of
complex fields $\varphi$ \cite{DeNicola2019,DeNicola2019long}. Focusing upon the case where $a = b$ in
(\ref{eq:heisenberg}), we can diagonalize the $N\times N$ matrix
$(\mathbf{J}^{aa})^{-1}$, for a given $a$. Explicitly, we may write
$(\mathbf{J}^{aa})^{-1} = \mathbf{V}^{a} \mathbf{D}^{a}
(\mathbf{V}^a)^{-1}$ where $\mathbf{D}^{a}$ is a diagonal matrix and
$\mathbf{V}^{a}$ is an eigenvector matrix, where $a$ labels the
matrices and not their components. It is also convenient to introduce
the white noises \cite{Ringel2013} $\phi^a_i(t) =\sum_j
(\mathbf{D}^{a})_{ii}^{1/2} ({\mathbf V}^{a})^{-1}_{ij}\varphi^a_j(t),$ which satisfy
\begin{align}
\langle \phi^a_i(t) \phi^b_j(t') \rangle = \delta^{ab}\delta_{ij}\delta(t-t'), ~~~ \langle \phi^a_i(t) \rangle =0.
\end{align} 
It follows from (\ref{eq:HStransf}) that the resulting dynamics are
described by a local stochastic Hamiltonian, $ \hat{H}^{s}_j(t) \!
\equiv \sum_a \! \Phi_j^a(t)\hat{S}^a_j ,\label{eq:hstocha}$ where
$\Phi_j^a(t) = \frac{-1}{\sqrt{i}}\sum_i (\mathbf{D}^{a})_{jj}^{-1/2}
\mathbf{V}^{a}_{ji}\phi^a_i(t) - h^a_j$. Letting $\langle ... \rangle_{\phi} =
\int D\mu(\phi) ... $ denote the average with respect to the Gaussian
measure, the time-evolution operator can be expressed as an average
over stochastic evolution operators. Explicitly, $\hat{U}(t) =
\Big\langle \hat{U}^{s} (t) \Big\rangle_{\phi}$ where $\hat{U}^{s} (t)
= \prod_j\hat{U}^{s}_j(t)$ and $\hat{U}^s_j(t) = \mathbb{T} e^{-i
  \int_0^t dt' \hat{H}^s_j (t')}$; here we write $\hat{U}(t)\equiv
\hat{U}(t,0)$ for brevity. Since $\hat{U}^{s}_j(t)$ has an
exponent that is linear in the $\hat{S}_j^a$  with complex
coefficients, $\hat{U}_j^s(t)$ is an element of $SL(2,\mathbb{C})$. As
such it can be re-expressed as a product of exponentials without 
time-ordering, \textit{via} a disentanglement transformation
\cite{Wei1963,Kolokolov1986,Ringel2013}.
Using the Gauss parametrization \cite{Klimov2009,Ringel2013}
\begin{equation} \hat{U}^{s}_j(t) = \mathbb{T} e^{-i \int_{0}^{t} dt'  \Phi^a_{j}(t') \hat{S}^a_{j} } \equiv e^{\xi^+_j(t) \hat{S}^+_j} e^{\xi^z_j(t) \hat{S}^z_j} e^{\xi^-_j(t) \hat{S}^-_j} \label{eq:gauss},\end{equation} 
where $\xi^{\pm,z}_j\in \mathbb{C}$ are referred to as disentangling variables \cite{Ringel2013,DeNicola2019}. The disentanglement 
 is achieved independently on each site by solving the Schr\"{o}dinger equation $ i (\partial_t \hat{U}^{s}_j) \hat{U}^{s~-1}_j  = \sum_a \Phi_j^a(t) \hat{S}_j^a $. This yields \cite{Ringel2013}
 \begin{subequations}
   \label{eq:SDEs}
\begin{align} & i \dot{\xi}^+_j = \Phi^+_j + \Phi^z_j \xi^+_j - \Phi^-_j \xi^{+^2}_j       , \label{eq:plus} 
\\ & i \dot{\xi}^z_j = \Phi^z_j-2 \Phi^-_j \xi^+_j, \label{eq:zequat} \\ &                                                       
i \dot{\xi}^-_j = \Phi^-_j e^{\xi^z_j}.  \label{eq:mininit}  \end{align}
\end{subequations}
where $ \Phi^{\pm}_j \equiv \frac{1}{2} (\Phi^x_j \mp i \Phi^y_j) $.  
The equations (\ref{eq:SDEs}) are SDEs for the $\xi$-variables due to
the stochastic fields $\Phi_j^a(t)$ \cite{Ringel2013}.  Quantum
observables, $\langle \hat{O}(t)\rangle$ are calculated as classical
averages over functions $f(\xi)$ of the $\xi$-variables, \textit{via} $\langle
\hat{O} \rangle = \langle f(\xi) \rangle_{\phi}$. In order to evolve
the $\xi$-variables forward in time, we solve the SDEs (\ref{eq:SDEs})
using a stochastic Heun predictor-corrector method in the Stratonovich formalism
\cite{Ruemelin1982,Kloden1992}. We find that this is capable of maintaining
accuracy with larger time-steps than the Euler-Maruyama scheme used
previously \cite{DeNicola2019,DeNicola2019long}, thereby reducing the
computational cost.

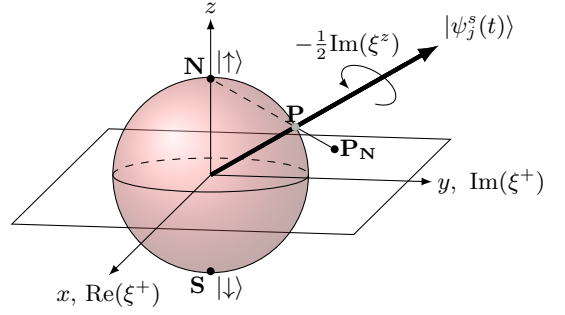
\begin{figure}[t]
\setlength{\abovecaptionskip}{-20pt}
\vspace{0cm}
\centering
\begin{tikzpicture} 
\newcommand{\AxisRotator}[1][rotate=0]{%
    \tikz [x=0.25cm,y=0.60cm,line width=.2ex,-stealth,#1] \draw (0,0) arc (-150:150:1 and 1);%
}
\def\angPhi{-100} 
\def\angBeta{30} 
\def\angBetamid{30} 
\def\Dobextra{3.7}
\def\Dob{3.5} 
\def\Dobmid{2.6} 
\def\Dobmidone{3.0}
\def\Dobmidtwo{2.2}  
\def\R{1.3} 
\def\angEl{10} 
\def\angAz{-100} 
\pgfmathsetmacro\H{\R*cos(\angEl)} 
\tikzset{xyplane/.style={cm={cos(\angAz),sin(\angAz)*sin(\angEl),-sin(\angAz),
                              cos(\angAz)*sin(\angEl),(0,0)}}}
\LongitudePlane[xzplane]{\angEl}{\angAz}
\LongitudePlane[pzplane]{\angEl}{\angPhi}
\LatitudePlane[equator]{\angEl}{0}
\draw[xyplane] (-2.5*\R,-1.5*\R) rectangle (3.2*\R,2.05*\R);
\begin{pgfinterruptboundingbox}
\shade[ball color = red!60, opacity = 0.4] (0,0) circle (\R); 
\draw (0,0) circle (\R);
\draw[ball color = white, opacity = 0] (0,0) circle (\Dob); 
 \end{pgfinterruptboundingbox}
\draw[ball color = white, opacity = 0] (0,0) circle (\Dobmid); 
\coordinate (O) at (0,0);
\coordinate[mark coordinate] (N) at (0,\H);
\coordinate[mark coordinate] (S) at (0,-\H);
\path[pzplane] (\angBeta:\R) coordinate[mark coordinate] (P);
\path[pzplane] (\angBeta:\Dob) coordinate (V);
\path[pzplane] (\angBetamid:\Dobmid) coordinate (Mid);
\path[pzplane] (\angBetamid:\Dobmidone) coordinate (Midone);
\path[pzplane] (\angBetamid:\Dobmidtwo) coordinate (Midtwo);
\path[pzplane] (\R,0) coordinate (PE);
\path[xzplane] (\R,0) coordinate (XE);
\path (PE) ++(0,0) coordinate (Paux); 
\coordinate[mark coordinate] (Phat) at (intersection cs: first line={(N)--(P)},
                                       second line={(S)--(Paux)});
\DrawLatitudeCircle[\R]{0} 
\draw[xyplane,<->] (6.0*\R,0) node[below] {$x,$~Re$(\xi^+)$} -- (0,0) -- (0,2.3*\R)
    node[right] {$y,~$Im$(\xi^+)$};
\draw[->] (0,0) -- (0,1.6*\R) node[above] {$z$};
\draw[dashed] (P) -- (N) node[above left] {$\mathbf{N}$}  node[above right] {$\ket{\uparrow}$};
\draw (P) -- (Phat) node[right] {$\mathbf{P_N}$};
\draw (S)  node[below left] {$\mathbf{S}$};
\draw (S)  node[below right] {$\ket{\downarrow}$};
\draw[ultra thick] (O) -- (P) node[above] {$\mathbf{P}$};
\draw[ultra thick, ->] (O) -- (V) node[fill=white, above right] {$\ket{\psi^s_j(t)}$};
\draw[ultra thick, ->] (P) -- (V)  node [midway] {\AxisRotator[thin,x=0.2cm,y=0.4cm,->,rotate=60]};
\draw[ultra thick] (P) -- (Midone)  node [above left]{\small{$-\frac{1}{2}\text{Im}(\xi^z)$}};
\path[pzplane] (\angBeta:\R) coordinate[mark coordinate, color = white, opacity = 0.7] (P);
\end{tikzpicture}
\caption{Projection of the Bloch sphere for an un-normalized quantum spin onto the complex plane, parametrized by $\xi^+$. The point $\bold{P}$ on the unit sphere is projected onto the point $\bold{P}_\bold{N}$ \textit{via} the North pole, $\bold{N}$. The point $\xi^+ = 0$ corresponds to spin-down $\ket{\downarrow}$, whilst $|\xi^+|\rightarrow \infty$ corresponds to spin-up $\ket{\uparrow}$. Potential divergences associated with $|\xi^+|\rightarrow \infty$ can be avoided \textit{via} a two-patch parametrization: the upper (lower) hemisphere is parametrized by projection from the South (North) pole. A mapping between the two patches is performed at the equator.}
 \label{fig:Riemann}
 \setlength{\abovecaptionskip}{10pt}
\end{figure}

{\em Parametrization.}--- To gain some intuition into the dynamics of
the SDEs (\ref{eq:SDEs}), it is instructive to consider the parametrization (\ref{eq:gauss}) in more detail. The stochastic evolution operator $\hat{U}^s_j(t)$ has a particularly simple form when acting on spin-down states \cite{DeNicola2019,DeNicola2019long}, due to the explicit form of the parametrization (\ref{eq:gauss}):
\begin{align}  \ket{\psi^s_j(t)} = \hat{U}^s_j(t) \ket{\downarrow}= \big( \ket{\downarrow}  + \xi^+_j(t)\ket{\uparrow} \big)e^{-\frac{\xi^z_j(t)}{2}},\end{align}
where the variable $\xi^-_j(t)$ drops out. Any stochastic state $\ket{\psi^s(t)}= \prod_j \ket{\psi^s_j(t)}$ can be parametrized in this way by introducing a preparation stage in which the initial state, $\ket{\psi_j^s(0)},$ is obtained as a rotation from a spin-down state $\ket{\downarrow}$; see Supplementary Material.

For a normalized spin state, spin-down $ \ket{\downarrow} $
corresponds to $\xi^+_j(t) = 0$ and spin-up $\ket{\uparrow} $
corresponds to $|\xi^+_j(t)| \rightarrow \infty$; this is a
stereographic projection of the Bloch sphere, \textit{via} the North pole, as
shown in Fig. \ref{fig:Riemann}. The complex parameter $\xi^z_j(t)$
determines the amplitude and phase of the spin state.  Divergences in
the SDEs (\ref{eq:SDEs}) \cite{DeNicola2019,DeNicola2019long} corresponding to $|\xi^+_j(t)| \rightarrow
\infty$ can be avoided by a
two-patch parametrization of the Bloch sphere by projecting from the
South pole for states in the upper hemisphere. This can be implemented
by the change of variables
\begin{subequations}
\begin{align}
& \xi^+_j(t)  \rightarrow \bar{\xi}^+_j(t) \equiv 1/\xi^+_j(t) ,  \label{eq:plusmap} \\  
& \xi^z_j(t) \rightarrow \bar{\xi}^z_j(t) \equiv \xi^z_j(t) - 2\ln(\xi^+_j(t)) \label{eq:zmap}, 
\end{align}
\end{subequations}
whenever the spins cross the equator. The corresponding SDEs for the new
coordinates are given in the Supplementary
Material. This approach for avoiding divergences in SDEs has also been
used in \cite{Ng2013}. We will use this two-patch approach throughout
the manuscript.  For simplicity, we focus on the nearest neighbor spin-1/2 ferromagnetic
quantum Ising model
\begin{equation} \hat{H}_I =   - \frac{J}{2} \sum_{\langle ij \rangle}  \hat{S}^z_i \hat{S} ^z_{j}  - \Gamma \sum_{j=1}  \hat{S}^x_j, \label{eq:Ising} \end{equation}
where we impose periodic boundary conditions. Throughout this paper we set $J = 1$ in the simulations. 
\begin{figure}[t!]
\setlength{\abovecaptionskip}{-3pt}
\includegraphics[width = 8.75cm]{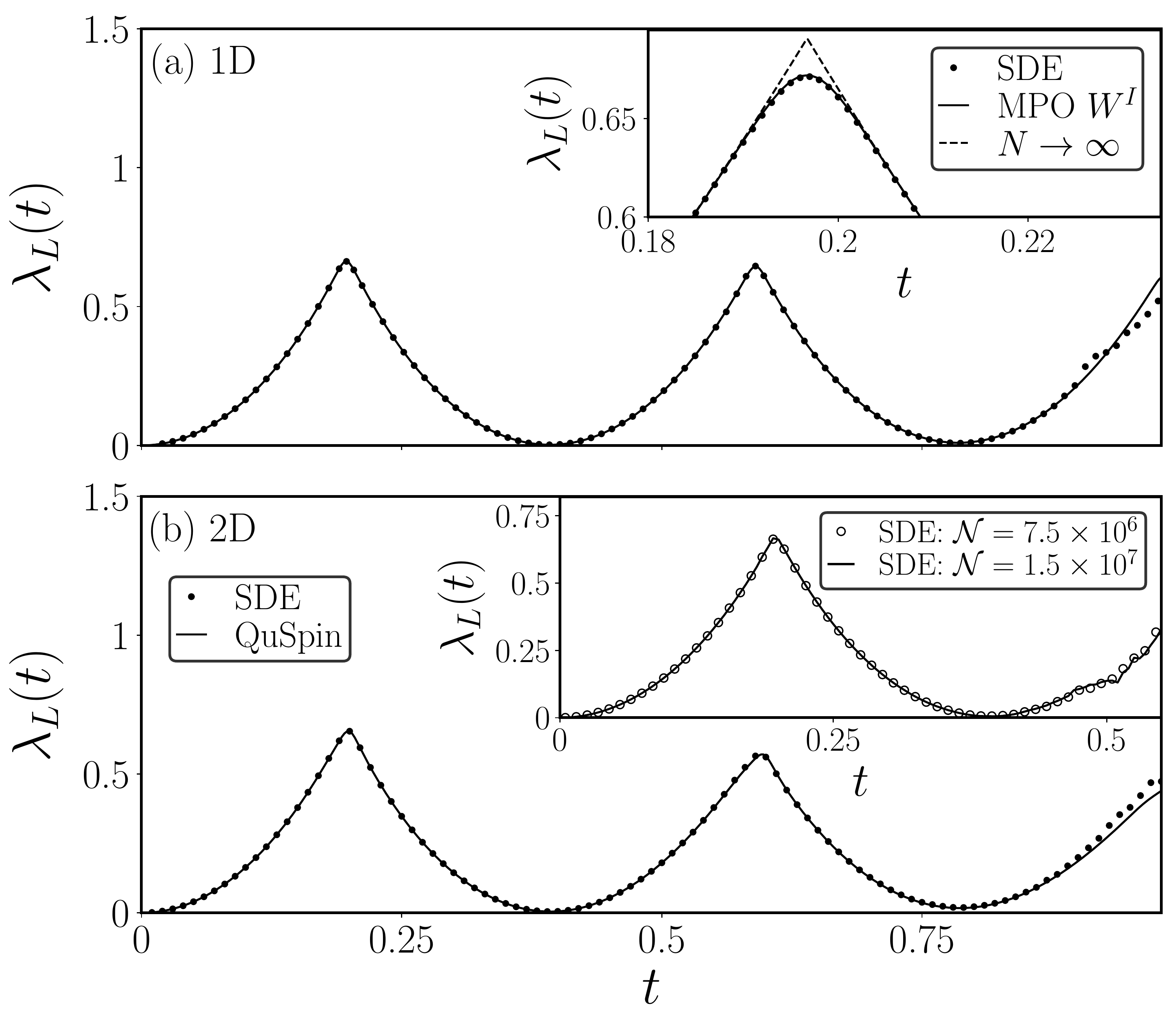}
\caption{Loschmidt rate function, $\lambda_L(t)$, following a quantum
  quench in the 1D and 2D quantum Ising model. We start
  in the ground state $\ket{\psi(0)} = \frac{1}{\sqrt{2}} (
  \ket {\Downarrow} +\ket {\Uparrow})$ for $\Gamma=0$ and quench to
  $\Gamma=8J$. (a) 1D case with $N=50$ spins. The results obtained from the SDEs (dots) are in
  excellent agreement with those obtained \textit{via} the MPO $W^{I}$ method (solid line). Deviations
  occur for $t\gtrsim 1/J$ as the stochastic fluctuations
  become harder to sample. 
  The inset shows a zoomed in portion of the first Loschmidt peak for
  $N=75$ spins, demonstrating similar agreement with MPO $W^{I}$. For
  comparison, we show the exact results of Heyl {\em et al.}
  \cite{Heyl} in the thermodynamic limit (dashed line). It is readily
  seen that the rounding of the Loschmidt peak is a finite-size effect.
  In both figures we use a
  time-step of $dt = 10^{-2}$, except in the vicinity of the 
  peaks, where $dt = 10^{-3}$ is used. The results are obtained by
  averaging over $ 10^7$ stochastic samples. (b) 2D case for a $5 \times 5$ lattice using $1.5 \times 10^7$ samples. The results are in agreement with QuSpin \cite{Weinberg2019} (solid line). The inset shows results for a $7 \times 7$ lattice, which cannot be obtained using QuSpin. Convergence is checked by changing the number of samples.}
\label{fig:rate}
\setlength{\abovecaptionskip}{10pt}
\end{figure}

{\em Loschmidt Amplitude.}--- As discussed in \cite{DeNicola2019,DeNicola2019long}, one
of the simplest quantities to examine in the stochastic approach is
the Loschmidt amplitude, $A(t) \equiv 
\bra{\psi(0)}\hat{U}(t)\ket{\psi(0)}$. The corresponding rate function
$\lambda_L(t) \equiv -\frac{1}{N}\ln|A(t)|^2$ plays a similar role
to the equilibrium free energy density: as $N\rightarrow\infty$ it
exhibits non-analytic peaks at DQPTs \cite{Heyl}. We first consider the one-dimensional case. In order to compare
to results obtained in the thermodynamic limit \cite{Heyl} it is convenient to
evolve from the ground state $\ket{\psi(0)} =
\frac{1}{\sqrt{2}} ( \ket {\Downarrow} + \ket {\Uparrow})$ at
$\Gamma=0$. Here $\ket {\Uparrow}$ and $\ket {\Downarrow}$ correspond
to the states with all the spins pointing up and down respectively.
Time-evolving these separately
using the SDEs (\ref{eq:SDEs}) one obtains:
\begin{align} A(t) &  = \frac{1}{\sqrt{2}}\left( \bra{\psi(0)}\hat{U}^s(t)\ket{\Downarrow}_\phi  + \bra{\psi(0)} \hat{U}^s(t)\ket{\Uparrow}_\phi \right),
\label{eq:loschmidtvar}
 \end{align}
where
$
 \bra{\psi(0)}\hat U^s(t)\ket{\Downarrow}  = \frac{1}{\sqrt{2}}\prod_j (1+\xi^+_j(t)) e^{\frac{-\xi^z_j(t)}{2}}, 
 $ and $\bra{\psi(0)}\hat U^s(t)\ket{\Uparrow}$ is obtained by
 $\xi^a\rightarrow {\bar\xi}^a$. The SDEs are solved with the initial
 conditions $\xi^a(0)=0$ and $\bar\xi^a(0)=0$ respectively.
 The results for $\lambda_L(t)$
 corresponding to time-evolution with $\Gamma=8J$ are shown in
 Fig. \ref{fig:rate}(a), for a 1D system with $N = 50$ spins. The results go
 beyond what is achievable using Exact Diagonalization (ED) and are in
 excellent agreement with those obtained \textit{via} the MPO $W^{I}$ method \cite{Zaletel2015}, implemented using ITensor \cite{ITENSOR}. Deviations are observed for $t\gtrsim
 1/J$ as the stochastic fluctuations become harder to sample.
 The inset shows the first Loschmidt peak for
 $N = 75$, which again demonstrates excellent agreement. For comparison we
 display exact results obtained in the thermodynamic limit, $N \rightarrow
 \infty$ \cite{Heyl}. Although the finite-size effects are stronger in
 the vicinity of the peak, the SDE and  MPO results 
 remain coincident for all of the system sizes considered. Results for the same quench on a $5 \times 5$ lattice in 2D are shown in Fig. \ref{fig:rate}(b) (dots), and are verified against those obtained using QuSpin's time-evolution solver \cite{Weinberg2019} (solid line). The inset shows the first Loschmidt peak for a $7 \times 7$ lattice, which goes beyond what we can readily verify using other techniques. We check for convergence near the peak by doubling the number of samples, $\mathcal{N}$, and noting that the results change by less than $0.5\%$ in this region.

\begin{figure}[t!]
\setlength{\abovecaptionskip}{2pt}
\includegraphics[width = 8.4cm]{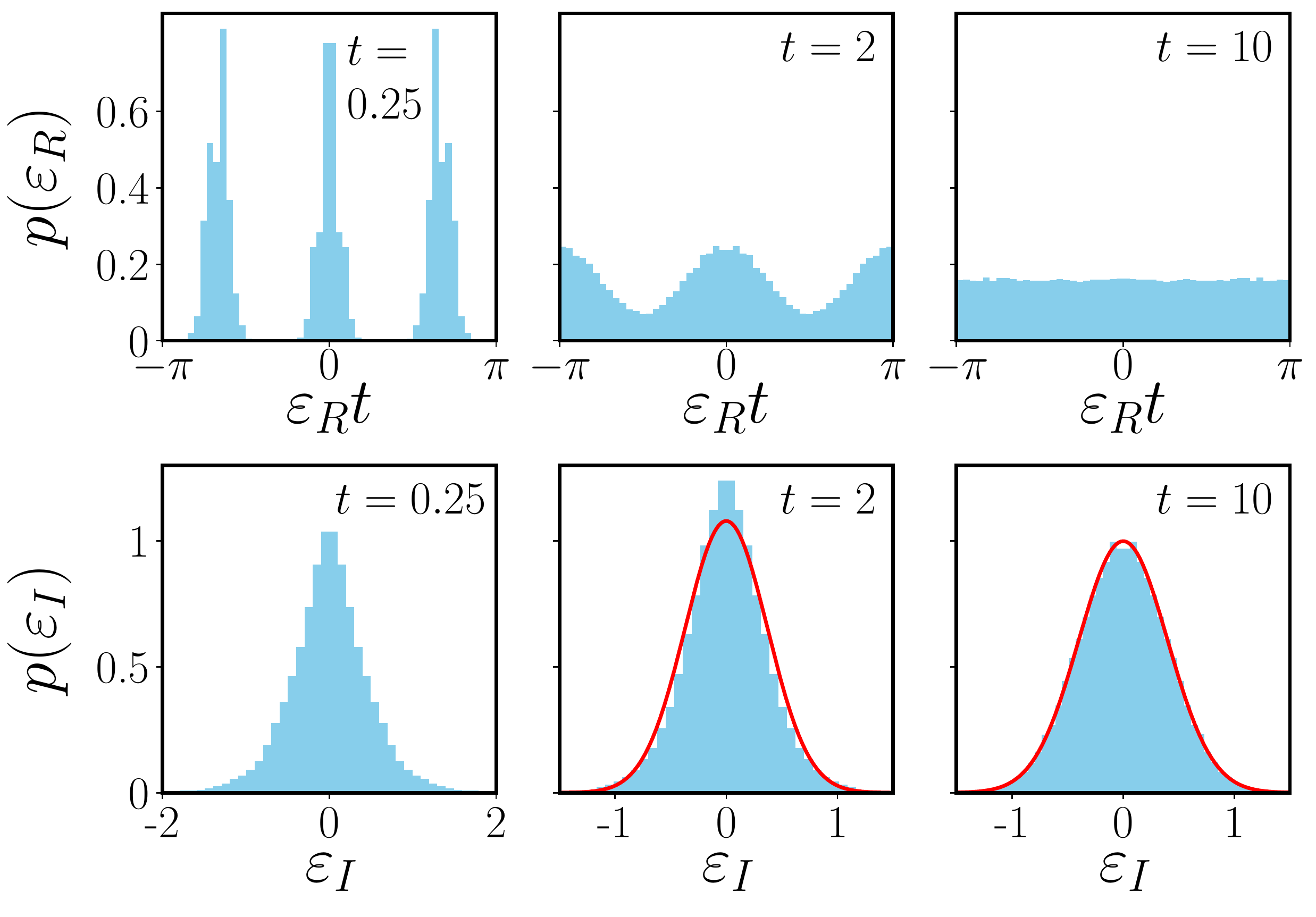}
\caption{Normalized eigenvalue distributions of the effective
  Hamiltonian $\hat{H}_{\text{eff}}$ for the 1D quantum Ising model
  with $\Gamma = 8J$ and $N=10$, obtained \textit{via} the SDEs with $dt = 10^{-4}$. The upper and lower panels show the
  real and imaginary parts, $\varepsilon_R$ and $\varepsilon_I$
  respectively, at times $t=0.25,2,10$; the abscissa in the upper
  plots is scaled by time.  The results correspond to a small number of samples, $\mathcal{N} =100$, to illustrate the non-Hermitian character of the stochastic representation. The distribution of
  $\varepsilon_R$ is approximately uniform at late times whereas the distribution of
  $\varepsilon_I$ is approximately normal, as shown by the solid (red)
  lines.
}
\label{fig:typ} 
\setlength{\abovecaptionskip}{10pt}
\end{figure}

{\em Growth of Fluctuations.}--- To quantify the role of stochastic
fluctuations it is instructive to consider the spectrum of an
effective Hamiltonian, $\hat{H}_{\text{eff}}(t)$, defined by
\begin{align} 
\hat{U}^s(t) = \prod_j e^{\xi^+_j(t) \hat{S}^+_j} e^{\xi^z_j(t) \hat{S}^z_j} e^{\xi^-_j(t) \hat{S}^-_j} \equiv e^{-i \hat{H}_{\text{eff}} t} ,\label{eq:effHam}\end{align} 
in analogy to Floquet systems \cite{Rahav2003}. Since $\hat{U}^s(t)$
is non-unitary, the eigenvalues of $\hat{H}_{\text{eff}}$,
$\varepsilon = \varepsilon_R + i \varepsilon_I$, are generically
complex.  The spectrum of $\hat{H}_{\text{eff}}(t)$ can be calculated
directly from (\ref{eq:effHam}) by time-evolving the SDEs to the time
of interest. This can also be obtained by noting that $\hat{U}^s(t)$ can
be calculated directly as a product of random matrices,
$\hat{U}^s(t)=\hat{U}^s(t,t-\delta)\hat{U}^s(t-
\delta,t-2\delta)...\hat{U}^s(\delta,0)$, by time-slicing into small
intervals of size $\delta$, without the disentangling
transformation.
In Fig. \ref{fig:typ} we show the time-evolution of the eigenvalue
distribution of $\hat H_{\rm eff}$, for $100$ stochastic
samples with $\Gamma = 8J$ and $N = 10$ in 1D. It can be seen that the
distribution of $\varepsilon_R$ is uniform at late times, whilst that of $\varepsilon_I$ is well approximated by a normal distribution.
In the Supplementary Material we show that the variance of the distribution of $\varepsilon_I$, denoted by $\sigma^2(t) =  \langle\varepsilon_I^2\rangle - \langle\varepsilon_I\rangle^2$,  exhibits damped oscillations as a function
of time, with extrema that occur in proximity to those in the
time-dependent magnetization.
In general, the presence of the positive imaginary eigenvalues results
in the growth of the norm of individual stochastic states over
time.  Due to the effect of finite numerical sampling,
this leads to the growth of the norm of the overall quantum state, and
to departures for physical observables. As we will see below, this
  can be partially compensated by rescaling by the norm.

\begin{figure}[t!]
\setlength{\abovecaptionskip}{2pt}
\includegraphics[width = 8.75cm]{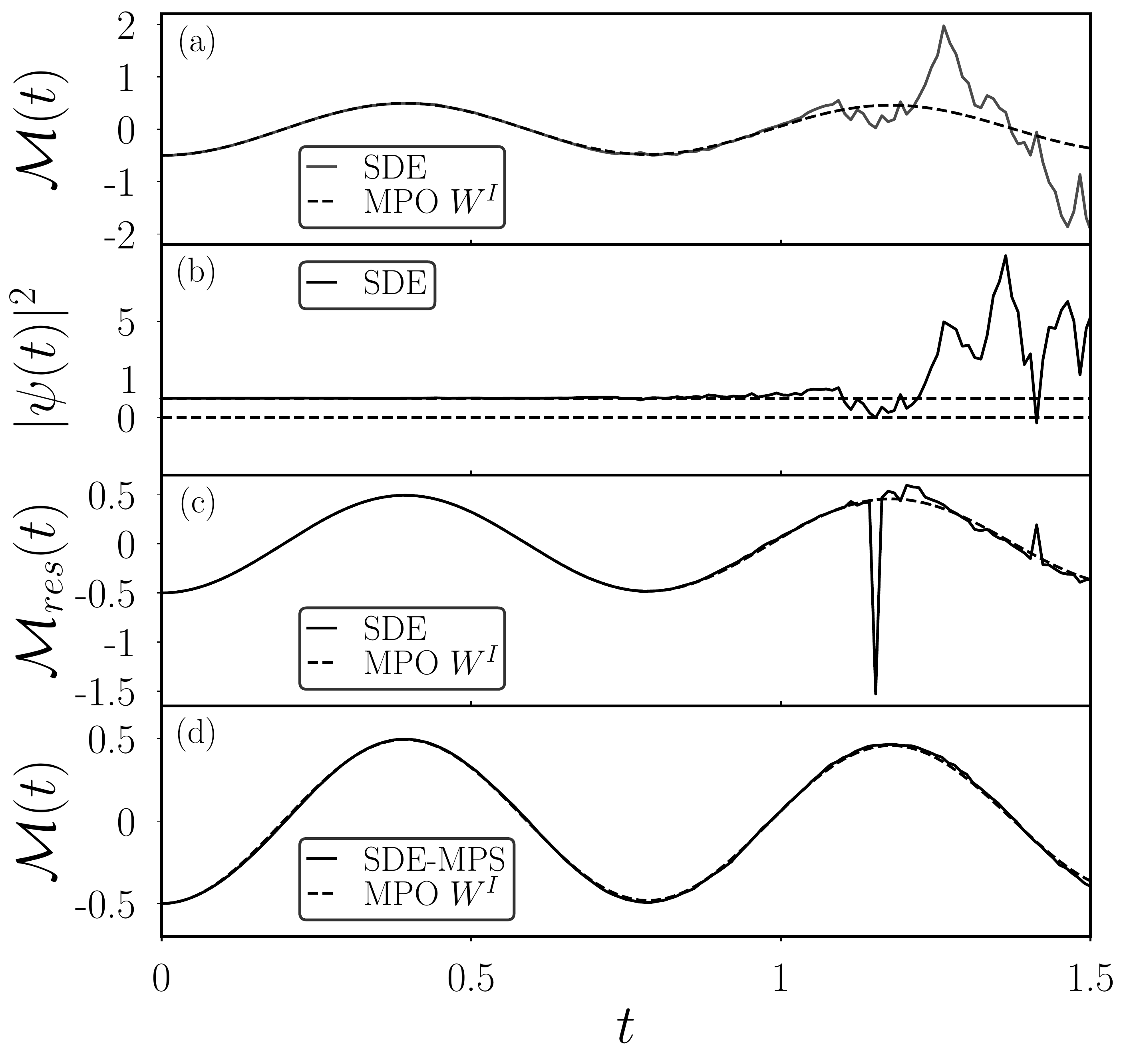}
\caption{(a) Time-dependent magnetization $\mathcal{M}(t)$ following a
  quantum quench in the 1D quantum Ising model from the
  fully-polarized initial state $\ket{\Downarrow}$ to $\Gamma=8J$ with
  $N=25$ spins. The results obtained from the SDEs (solid line) with
  $2.5 \times 10^6$ samples are in agreement with
   MPO $W^{I}$ (dashed line) until $t\sim  1/J$. (b) Time-evolution of
  the norm of the quantum state $|\psi(t)|^2$ following the quench in
  (a). The norm departs from unity when the stochastic fluctuations
  become significant. (c) Time-evolution of the rescaled magnetization
  $ \mathcal{M}_{res}(t)$ showing better agreement with  MPO $W^{I}$. It can be seen that fluctuations in
  $\mathcal{M}_{res}(t)$ occur whenever $|\psi(t)|^2$ is close to zero. (d)
  Time-evolution of ${\mathcal M}(t)$ using a hybrid stochastic-MPS
  approach for $N=50$ spins, with $50,000$ samples and a maximum
  bond dimension of $D_i = 20$. The results are in good agreement
  with  MPO $W^{I}$  (dashed line).}
  \label{fig:h8mps}
  \setlength{\abovecaptionskip}{0pt}
\end{figure}

{\em Magnetization Dynamics.}--- As discussed in
\cite{DeNicola2019,DeNicola2019long}, time-dependent physical
observables can be obtained by using two Hubbard--Stratonovich
transformations to decouple the forwards and backwards evolution
operators:
\begin{equation} 
\langle  \hat{O}(t)  \rangle = \bra{\psi(0)} \hat{U}^{s \dagger}(\tilde{\phi}) \hat{O} \hat{U}^{s} (\phi) \ket{\psi(0)}_{\phi,\tilde{\phi}},
\end{equation}
where $\phi$ and $\tilde\phi$ are independent noise variables. 
In this representation, the local magnetization is given by \cite{DeNicola2019}
\small
\begin{equation}
\langle  \hat{S}^z_j  \rangle =-\frac{1}{2} \Big\langle   e^{-\frac{1}{2}(\chi^z + \tilde{\chi}^{z*})}
  \Big(1 - \xi^+_j \tilde{\xi}^{+*}_j   \Big)   \prod_{i \neq {j}}  \Big(1+ \xi^+_i \tilde{\xi}^{+*}_i \Big)\Big\rangle_{\phi, \tilde{\phi}}, \label{eq:magnetizationform} 
\end{equation}
\normalsize where $\chi \equiv \sum_i \xi^z_i $, $\tilde{\chi} \equiv
\sum_i \tilde{\xi}^z_i$, and we implicitly take the real part; in
general, observables have imaginary parts which vanish in the limit of
infinite sampling \cite{Drummond1980,Barry2008,Ng2013}. In Fig.~\ref{fig:h8mps}(a) we show results for the time-dependent
magnetization $\mathcal{M}(t) = \frac{1}{N} \sum_j \langle
\hat{S}^z_j(t) \rangle,$ following a quantum quench from the
fully-polarized initial state $\ket{\Downarrow}$ to $\Gamma=8J$
for a 1D system with $N =25$ spins. The results are in good agreement with
 MPO calculations until times $t\gtrsim 1/J$ when stochastic fluctuations become
large. In Fig.~\ref{fig:h8mps}(b) we show results for the norm of
the time-evolving state as computed from the SDEs:
\begin{equation}
 |\psi(t)|^2 = \Big\langle   e^{-\frac{1}{2}(\chi^z + \tilde{\chi}^{z*})}  \prod_{i}  \Big(1+ \xi^+_i \tilde{\xi}^{+*}_i \Big)
 \Big\rangle_{\phi, \tilde{\phi}},\label{eq:normobs}
\end{equation}
where again, we take the real part. It is readily seen that the norm
departs from unity once the stochastic fluctuations become
significant. In Fig.~\ref{fig:h8mps}(c) we show results for the
re-scaled magnetization $\mathcal M_{res}(t)={\mathcal
  M}(t)/|\psi(t)|^2$ which provides much better agreement with the MPO results 
until later times. Fluctuations in $\mathcal M_{res}(t)$
still occur however, especially when the norm of the state is close to
zero. This clearly highlights the importance of normalization and stochastic
sampling in the computation of  observables.

{\em Matrix Product States.}---
Another approach to reducing fluctuations in observables is to decompose the time-evolving state into a matrix product state (MPS): 
\begin{align} \ket{\psi(t)} =\sum_{\substack{\sigma_1,..\sigma_n \\ d_1,..d_{n-1}}}A_{1,d_1}^{\sigma_1} A_{d_1,d_2}^{\sigma_2}....A_{d_{n-1},1}^{\sigma_n} \ket {\sigma_1 \sigma_2...\sigma_n}, 
 \end{align} 
where $\mathbf{A}^{\sigma_i}$ are matrices
with physical spin indices, $\sigma_i$, and auxiliary indices $d_i\in 1,\dots,D_i$, where $D_i$ are the bond dimensions; see
\cite{Schollwock2011} for an introduction. This reduces the number of
noise variables required since only a single Hubbard--Stratonovich transformation is
needed for time-evolution; see Supplementary
Material. One may also calculate the norm of the state using MPS
techniques, thereby eliminating fluctuations from the
stochastic sampling of $|\psi(t)|^2$. In Fig \ref{fig:h8mps}(d) we show the
results of the hybrid stochastic-MPS approach for a quantum quench in
a 1D system with $N = 50$ spins. The results are in
excellent agreement with the MPO approach, in spite of doubling 
the system size and reducing the number of stochastic samples.
A notable disadvantage of this hybrid approach is that one must
store the MPS state in memory at the expense of
the stochastic
parallelization. Nonetheless, the marriage of these approaches may
be useful for future developments.

{\em Conclusions.}--- 
In this work we have demonstrated that the
stochastic approach to non-equilibrium quantum spin systems can
address significantly larger systems than recently obtained,
in both one and two dimensions.
We have shown that the non-Hermitian character of the
representation leads to a growth of the norm of $\ket{\psi(t)}$, due to the effect of finite numerical sampling. However, this
can be compensated for by rescaling by the norm.  We have shown that the approach can be combined with a decomposition in terms of matrix
product states, for the calculation of time-dependent
observables. There are many directions for research, including
extensions to larger system sizes and later times, particularly in higher dimensions where few techniques are available.

{\em Acknowledgements.}--- We acknowledge helpful conversations with
S. De Nicola, B. Doyon, D. O'Dell and S. W\"{u}ster.  We also thank
J. Morley for assistance with the implementation of MPS algorithms.
SEB is supported by the EPSRC CDT in
Cross-Disciplinary Approaches to Non-Equilibrium Systems (CANES) \textit{via}
grant number EP/L015854/1. We are grateful to the UK Materials and
Molecular Modelling Hub for computational resources, which is
partially funded by EPSRC (EP/P020194/1). The MPO calculations were performed using the ITensor Library \cite{ITENSOR}. MJB acknowledges the
support of the ICTS (Bengaluru) during the program on
Non-Hermitian Physics PHHQP XVIII. AGG acknowledges EPSRC grant 	EP/P013449/1.

\vspace{-0.6cm}

\bibliographystyle{apsrev4-1}

\onecolumngrid 
\newpage
\section{Supplementary Material }

\vspace{0.15cm}

\section{I. Parametrization} \label{sec:stateprep}
As discussed in the main text, we may eliminate divergent trajectories from the SDEs (5) by a suitable parametrization of the stochastic time-evolution operator, $\hat{U}^s_j(t)$. Adopting the Gauss parametrization of SL(2,$\mathbb{C}$) \cite{Ringel2013}
\begin{align} \hat{U}^{s}_j(t) \equiv e^{\xi^+_j(t) \hat{S}^+_j} e^{\xi^z_j(t) \hat{S}^z_j} e^{\xi^-_j(t) \hat{S}^-_j} \label{eq:gauss}.\end{align}
For spin-$1/2$ systems this can be represented in matrix form as
\begin{equation}
\hat{U}^{s}_j(t)  = \begin{pmatrix}
e^{\frac{1}{2} \xi^z_j} + \xi^+_j \xi^-_j e^{-\frac{1}{2} \xi^z_j} & \xi^-_j e^{-\frac{1}{2}\xi^z_j} \\  \xi^+_j e^{-\frac{1}{2} \xi^z_j} &  e^{-\frac{1}{2}\xi^z_j} \label{eq:gaussmatrix}
\end{pmatrix},
\end{equation}
where $\xi^{\pm,z}_j \in \mathbb{C},$ and the initial conditions $\xi^{\pm,z}_j(0) = 0$ ensure that $\hat{U}^s_j(0) =\mathbb{1}$ is the identity operator. In general, this corresponds to a representation of the group SL(2,$\mathbb{C}$) with three complex parameters. The action of $\hat{U}^s_j(t)$ on a generic initial state $\ket{\psi_j(0)} = a\ket{\downarrow} + b \ket{\uparrow}$, with $a,b \in \mathbb{C}$ yields
\begin{equation}
\ket{\psi^s_j(t)} =   e^{-\frac{1}{2}\xi^z_j(t)} \Big[ \big( a + b \xi^-_j(t) \big) \ket{\downarrow}  +  \big( a \xi^+_j(t)  +  b e^{\xi^z_j(t)} +  b \xi^+_j(t) \xi^-_j(t) \big) \ket{\uparrow} \Big], \label{eq:genericstates}
\end{equation}
Although (\ref{eq:genericstates}) is formally exact, the parametrization contains some redundancy: an arbitrary un-normalized spin state can be represented by four parameters, including the overall phase. To see that a reduction is possible it is instructive to consider an initial spin-down state $\ket{\downarrow}$ corresponding to $a = 1$ and $b =0$. This yields
\begin{equation}
\ket{\psi^s_j(t)} = \hat{U}^{s}_j(t) \ket{\downarrow} =  e^{-\frac{1}{2}\xi^z_j(t)} \big( \ket{\downarrow}  +   \xi^+_j(t)   \ket{\uparrow} \big),
\label{eq:4dof}
\end{equation}
where the complex parameter $\xi^-_j(t)$ has dropped out. In this case, the divergence in the SDE (5a) corresponding to $|\xi_j^+|\rightarrow \infty$ is associated with an inability to parametrize the spin-up state $\ket{\uparrow}$, using a projective representation of the Bloch sphere. As discussed in the main text and in Section II below, this can be avoided by using a two-patch parametrization. Although these considerations apply only for an initially spin-down state, more general initial states can always be prepared by rotation from this state. Explicitly, we may introduce  a state-preparation protocol starting at $t = - \delta$, with $\delta > 0$, and evolving deterministically until $t =0$. In this approach, the time-evolution operator takes the form $\hat{U}^{s}_j(t,-\delta) = \mathbb{T} e^{-i \int_{-\delta}^{t}  \hat{H}^s_j(t')dt ' }$ where 
\begin{equation} 
 \hat{H}^s_j(t) = \begin{cases}
 \alpha_j^a \hat{S}^a_{j},~~~~~~ -\delta \leq t < 0;
   \\ \Phi^a_{j}(t') \hat{S}^a_{j},~~~~~~~~~~ t \geq 0,
\end{cases}  
\end{equation}
and the coefficients $\alpha_j^a$ specify the initial conditions according to $\ket{\psi^s_j(0)}  = \hat{U}^s_j(0,-\delta)\ket{\downarrow}$. In practice, this is equivalent to setting non-trivial initial conditions for the $\xi$-variables and evolving under the SDEs (5).
For example, the initial state $\ket{\psi_j(0)} = \frac{1}{\sqrt{2}} \big(\ket{\downarrow} + \ket{\uparrow} \big)$ corresponds to $\xi^+_j(0) = 1$ and $\xi^z_j(0) = \ln 2,$ as follows directly from (\ref{eq:4dof}). In this approach the trivial initial conditions correspond to $t = -\delta$, so that $\xi_j^+(-\delta) = \xi_j^-(-\delta)= \xi_j^z(-\delta) =0$ and $\hat{U}^s_j(-\delta,-\delta) = \mathbb{1}$.

In general, the time-evolution of an arbitrary product state is given by \begin{equation}\ket{\psi(t)} = \Big \langle \prod_j  e^{-\frac{1}{2}\xi^z_j(t)} \big( \ket{\downarrow}  +   \xi^+_j(t)   \ket{\uparrow} \big) \Big \rangle_{\phi}, \label{eq:prodstatexi}\end{equation}
where the initial conditions $\xi^a_j(0)$ specify the initial spin-orientation at each site. A generic superposition can be obtained by summing over (\ref{eq:prodstatexi}) with the appropriate initial conditions. 
Stochastic expressions for physical observables are readily obtained from the projective representation (\ref{eq:4dof}). For example, the Loschmidt amplitude to remain in the spin-down state is given by $A(t) = \langle e^{-\frac{1}{2}\sum_j \xi_j^z(t)} \rangle_{\phi}$, in agreement with 
(\ref{eq:loschmidtvar}) and \cite{DeNicola2019}. In a similar way, the quantum expectation value of the spin operator $\hat{\mathbf{S}}_j$ is given by \begin{align} \langle \hat{\mathbf{S}}_j(t) \rangle = \Big\langle \prod_i |\psi_i^s(t)|^2 \mathbf{n}_j(t) \Big\rangle_{\phi,\tilde{\phi}}, \label{eq:singlespin}\end{align}
where \begin{align} \mathbf{n}_j(t) = \frac{1}{2}\Bigg(\frac{2\text{Re}(\xi^+_j(t))}{1+ |\xi^+_j(t)|^2} ,\frac{-2\text{Im}(\xi^+_j(t))}{1+ |\xi^+_j(t)|^2}  ,\frac{-1 + |\xi^+_j(t)|^2}{1+ |\xi^+_j(t)|^2}  \Bigg) \label{eq:euclid} \end{align} corresponds to the position of a spin on the Bloch sphere. The factor of \begin{align}
|\psi_i^s(t)|^2 = e^{-\text{Re}(\xi^{z}_i(t))}(1 +  |\xi^+_i(t)|^2)  \label{eq:normstoch}
\end{align} is the norm of the state $\ket{\psi_i^s(t)}$. In writing (\ref{eq:singlespin}), (\ref{eq:euclid}) and (\ref{eq:normstoch}), it is implicit that $\xi^{a*}_j$ is an independent variable from $\xi^{a}_j$, which we denote by $\tilde{\xi}^{a*}_j = \xi^{a*}_j$ in the main text. The result (\ref{eq:singlespin}) is readily generalized to multipoint correlation functions. For example, 
\begin{align}
 \langle \hat{S}^a_j(t) \hat{S}^b_k(t)  \rangle = \Big\langle \prod_i |\psi_i^s(t)|^2 n_j^a(t) n_k^b(t) \Big\rangle_{\phi,\tilde{\phi}} .
\end{align}
The expressions for entangled states can be obtained by averaging over the initial conditions for $\xi(0)$ and $\tilde{\xi}(0)$.

\section{II. Eliminating divergent trajectories}\label{sec:divergences}
\begin{figure}[t]
\includegraphics[width = 11cm]{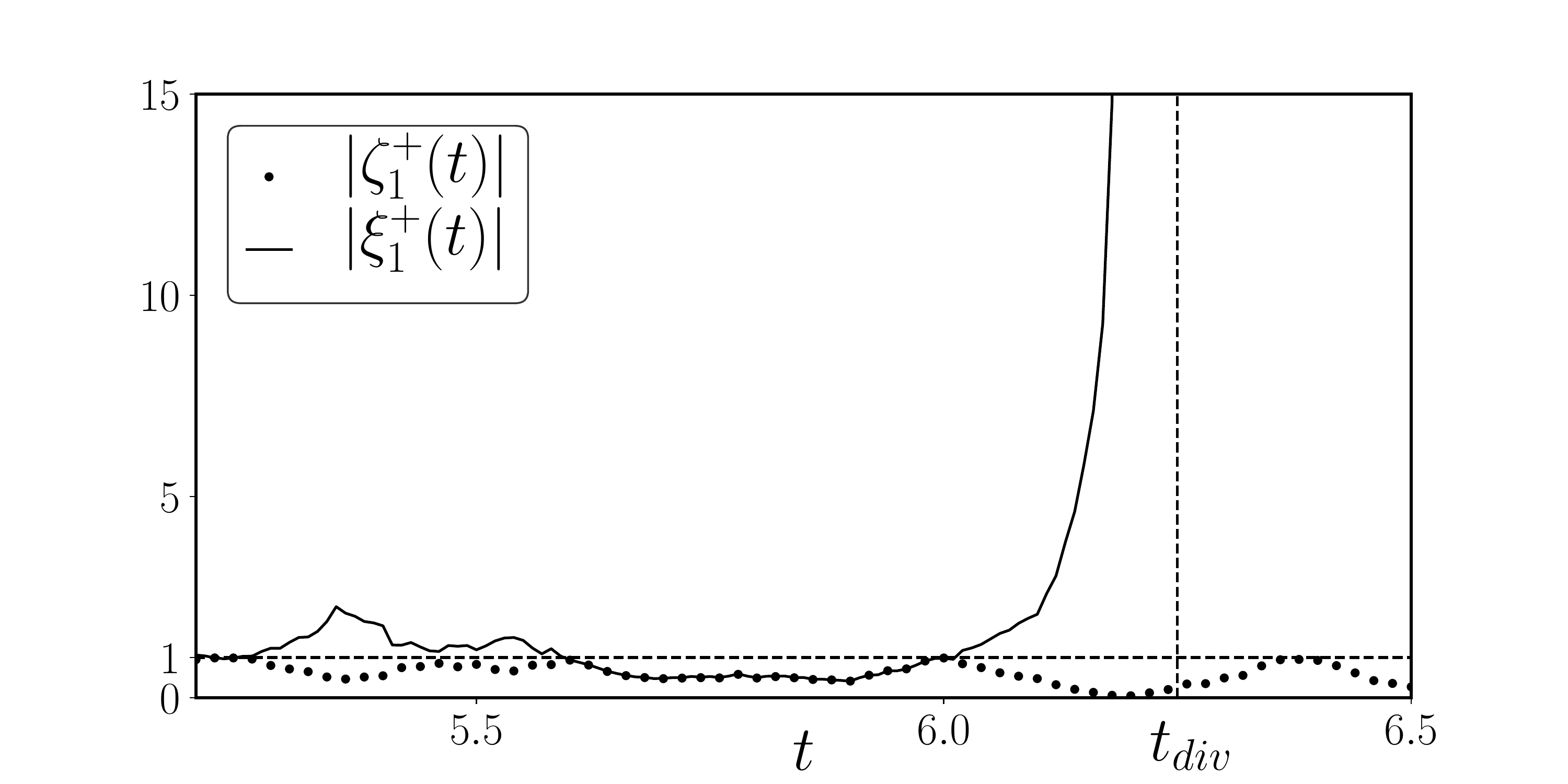} 
\caption{Time-evolution of a divergent trajectory $|\xi^+_1(t)|$ (solid line) following a quantum quench in the 1D quantum Ising model from the fully polarized state $\ket{\Downarrow}$ to $\Gamma = 8J$, with $N = 7$ spins. The diverging quantity is evaluated on the first site, and overflows at $t_{div} = 6.25 $. For comparison, the two-patch variable $|\zeta^+_1(t)|$ (dotted) does not diverge.}
\label{fig:divtraj}
\end{figure}

As discussed above and in the main text, the SDE (5a) exhibits divergences corresponding to $|\xi^+_j| \rightarrow \infty$. These can be avoided by two-patch parametrization of the Bloch sphere. To this end, it is convenient to define new variables $\bar{\xi}^a$ via the generalization of equation (\ref{eq:4dof}), where the roles of $\ket{\downarrow	}$ and $\ket{\uparrow}$ are interchanged:
\begin{equation} \ket{\psi^s_j(t)} = \Big( \bar{\xi}^+_j(t) \ket \downarrow  + \ket \uparrow  \Big) e^{-\frac{1}{2}\bar{\xi}^z_j(t)}. \label{eq:actup}\end{equation} 
Equating the coefficients of (\ref{eq:4dof}) and (\ref{eq:actup}), one obtains the identifications
\begin{equation} \bar{\xi}^+_j(t) = \frac{1}{\xi^+_j(t)}, ~ \bar{\xi}^z_j(t) =  \xi^z_j(t) -  2 \ln(\xi^+_j(t)). \label{eq:maps}\end{equation} 
This coordinate system is related to the original Gauss parametrization (\ref{eq:gauss}) by swapping the pole of projection from the North to the South pole. Performing this change of variables, the SDEs (5a) and (5b) become
\begin{subequations}
\begin{align} & i\dot{\bar{\xi}}^+_j = \Phi^-_j - \Phi^z_j \bar{\xi}^+_j - \Phi^+_j \bar{\xi}_j^{+2},       \label{eq:AGmin}      
\\ & i\dot{\bar{\xi}}^z_j = -\Phi^z_j -  2 \Phi^+_j\bar{\xi}^+_j .\end{align}
\label{eq:agausseq}
\end{subequations}
A convenient place to perform the change of variables (\ref{eq:maps}) is when the spins cross the equator, since the magnitude of $|\xi^+_j(t)| = |\bar{\xi}^+_j(t)| = 1$ at this point, and the numerical error associated with simulating the nonlinear term in (\ref{eq:plus}) is minimized. This approach is also used in \cite{Ng2013}.
In practice, the initial state will determine which parametrization is initialized on each site. 
Denoting \begin{equation} \zeta_j^+(t) = \begin{cases}
        \xi^+_j(t) ,~~~  \text{lower half-sphere with } |\xi_j^+|\in [0,1];  \\
      \bar{\xi}^+_j(t) ,~~~  \text{upper half-sphere with } |\bar{\xi}_j^+|\in [0,1),
    \end{cases} \label{eq:regions}
  \end{equation}
one may track the dynamics of this single variable over all times. In Fig. \ref{fig:divtraj} we plot the time-evolution of the single-patch variable $| \xi_1^{+}(t)|$ following a quantum quench in the 1D quantum Ising model. For the chosen parameters and the specific noise realization, it can be seen that this quantity diverges at time $t_{div} = 6.25$. As a result, $\dot{\xi}_1^{+}(t)$ overflows due to the $\xi_1^{+^2}(t)$ term in (\ref{eq:plus}), and numerical integration fails for this trajectory. In contrast, the two-patch variable $|\zeta^+_1(t)|$ remains finite, and we can evolve beyond $t_{div}$. This enables us to retain all the stochastic trajectories when computing the time-dependent magnetization, in contrast to previous work \cite{DeNicola2019,DeNicola2019long}.

\begin{figure}[t!]
\includegraphics[width = 11cm]{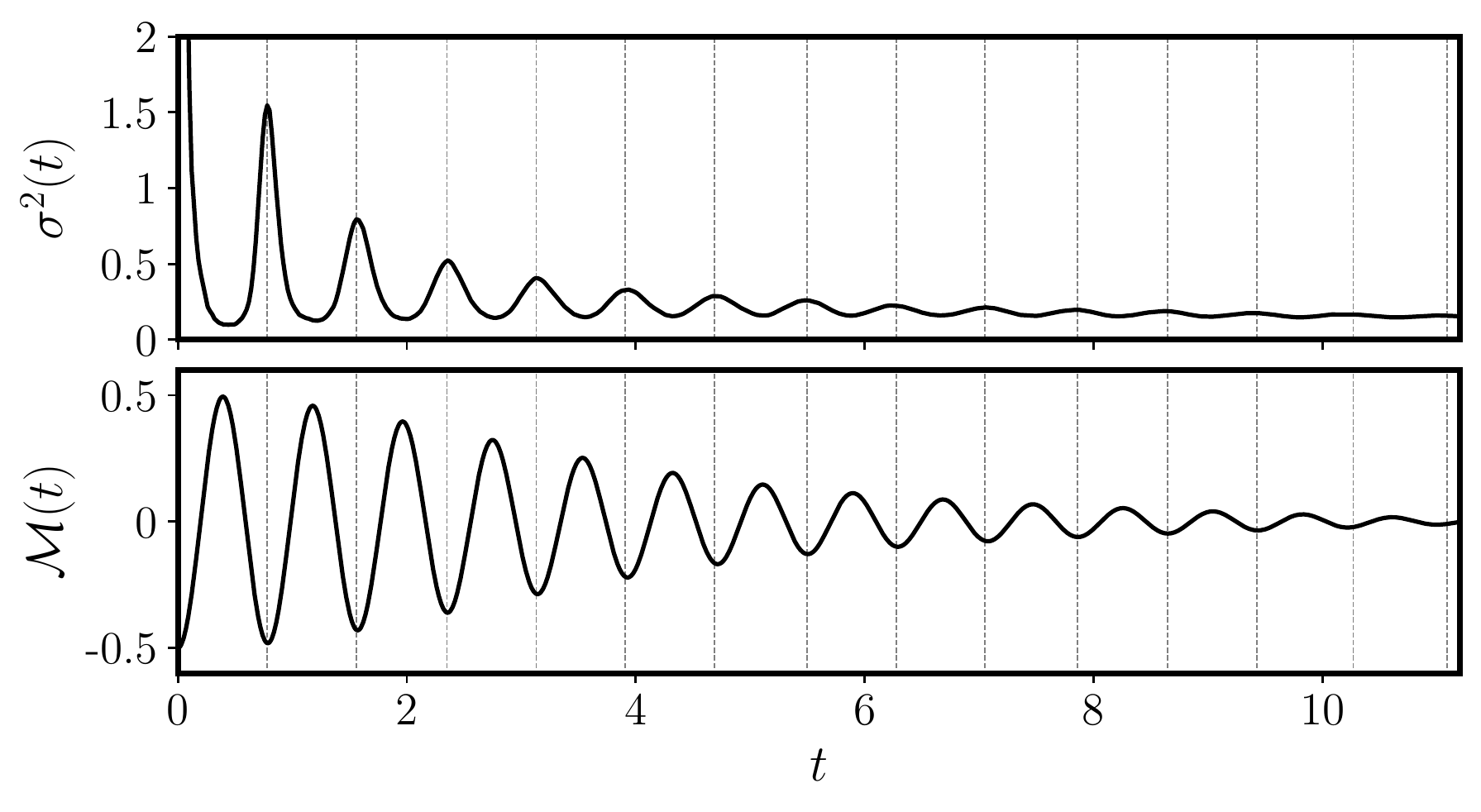}
\caption{ Normalized eigenvalue distributions of the effective
  Hamiltonian $\hat{H}_{\text{eff}}$ for the 1D quantum Ising model
  with $\Gamma = 8J$ and $N=10$. 
 The variance, $\sigma^2(t)$, of the distribution of $\varepsilon_I$ exhibits damped oscillations
    as a function of time. The extrema occur in proximity
    to the turning points in the magnetization, $\mathcal{M}(t)$, obtained via exact diagonalization, following a quench from the fully-polarized initial state $\ket{\Downarrow}$
to $\Gamma=8J$ for $N=10$ spins, as indicated by the vertical lines.}
\label{fig:vareps} 
\setlength{\abovecaptionskip}{10pt}
\end{figure}

To analyze the spectrum of $\hat{U}^s(t,0)$, or equivalently $\hat{H}_{\text{eff}} = \frac{i}{t} \ln \hat{U}^s(t,0)$ as illustrated in Fig. \ref{fig:typ}, the parameter $\xi^-_j$ is also required; this dropped out in (\ref{eq:4dof}). Under the transformation $(\xi^+_j,\xi^z_j)\rightarrow (\bar{\xi}^+_j,\bar{\xi}^z_j)$ the SDE (\ref{eq:mininit}) becomes
\begin{align} i\dot{\xi}^-_j = \Phi^-_j \frac{e^{\bar{\xi}^z_j}}{\bar{\xi}^{+2}}. \label{eq:tempsde} \end{align}
In order to ensure that (\ref{eq:tempsde}) is well-behaved as $\bar{\xi}^+_j \rightarrow 0$ it is convenient to make the change of variables 
\begin{align} \xi_j^- \rightarrow \bar{\xi}^-_j \equiv  \xi^-_j + \frac{e^{\bar{\xi}^z_j}}{\bar{\xi}^+_j} 
 .\label{eq:permutemapmin}\end{align}
The resulting SDE for $\bar{\xi}^-_j$ is given by
\begin{align}i\dot{\bar{\xi}}^-_j = -\Phi^+_j e^{\bar{\xi}^z_j},\label{eq:mappedmin} \end{align}
which mirrors (\ref{eq:mininit}) up to a sign change and $\Phi^-_j \rightarrow \Phi^+_j$. The maps (\ref{eq:maps}) and (\ref{eq:permutemapmin}) can now be conducted simultaneously to avoid the divergence associated with $\xi^+_j \rightarrow \infty$, allowing $\hat{U}^s(t,0)$ to be calculated at much later times. Eventually this strategy will break down if (\ref{eq:mininit}) or (\ref{eq:mappedmin}) cannot be integrated due to $e^{\xi^z_j} \rightarrow \infty$ or $e^{\bar\xi^z_j} \rightarrow \infty$ respectively. 
However, $\xi^-_j$ and $\bar{\xi}^-_j$ are not required for the time-dependent magnetization, this is not a limitation. The spectrum of $\hat{H}_{\text{eff}}$ (\ref{eq:effHam}) is calculated directly from trajectories by mapping between the $\xi^a_j$ and $\bar{\xi}^a_j$ variables in accordance with the prescription (\ref{eq:regions}). As discussed in the main text, damped oscillations of the variance of the imaginary eigenvalues of $\hat{H}_{\text{eff}}$ occur as a function of time; see Fig. \ref{fig:vareps}.

\section{III. Hybrid technique with matrix product states} \label{sec:MPS}
As discussed in the main text, the time-evolving quantum state evaluated within the stochastic approach can be represented as a matrix product state. This halves the number of noise variables required since only a single Hubbard--Stratonovich transformation is needed to evaluate $\ket{\psi(t)}= \hat{U}(t)\ket{\psi(0)}$. However, this comes at the cost of storing the state in memory, so the method is no longer fully parallelizable. Here we demonstrate how this representation is obtained. The quantum state is first written as the sample average of the stochastic state: 
\begin{align} \ket{\psi(t)} =  \ket{ \psi^{s}(t)}_{\phi} = \frac{1}{\mathcal{N}} \sum_{r = 1}^{\mathcal{N}} \prod_i^N \Big(\ket{\downarrow} + \xi^{+,r}_i  \ket{\uparrow}\Big)e^{-\frac{\xi^{z,r}_i}{2}}, \label{eq:mpsform}\end{align} 
where  $r = 1,...,\mathcal{N}$ is the sample index.
A matrix product state (MPS) is given by 
\begin{equation} \ket {\psi} =\sum_{\substack{\sigma_1,..,\sigma_N \\ d_1,..,d_{N-1}}}A_{1,d_1}^{\sigma_1} A_{d_1,d_2}^{\sigma_2}....A_{d_{N-1},1}^{\sigma_N} \ket {\sigma_1 \sigma_2...\sigma_N},\label{eq:matrixproductstateansatz} \end{equation}
where the matrices $\mathbf{A}^{\sigma_i}$ carry physical spin indices $\sigma_i$ and auxiliary indices $d_i\in 1,...,D_i$, where $D_i$ are the bond dimensions.  
To cast the state (\ref{eq:mpsform}) into MPS form, we first note that each configuration of indices $d_1, d_2, …$ in (\ref{eq:matrixproductstateansatz}) forms a product state that can be identified as one term in the sum (\ref{eq:mpsform}). We can identify a trivial, inefficient MPS representation of (\ref{eq:mpsform}) by taking a diagonal form for the MPS tensors. For example
\begin{equation} \mathbf{A}^{\downarrow} = \begin{pmatrix} 
e^{-\frac{\xi^{z,1}_i}{2}} & 0  & ....& 0 \\
0 & e^{-\frac{\xi^{z,2}_i}{2}} & ... & 0\\ 
 \vdots &  \vdots & \ddots & \vdots\\
0 & 0 &  ... & e^{-\frac{\xi^{z, \mathcal{N}}_i}{2}}\\ 
\end{pmatrix} ,\label{eq:mpselements}\end{equation} 
where the factor of $1/\mathcal{N}$ in (\ref{eq:mpsform}) can be absorbed into one of the $\mathbf{A}^{\sigma_i}$ matrices. The state $\ket{\psi}$ can be compressed to a lower bond-order, full rank MPS by using a sequence of singular value decompositions (SVDs).  In practice, we may perform this procedure in batches. The MPS tensors in each batch can be further combined and compressed by first collecting them into a block diagonal MPS tensor given by
 \begin{equation} \mathbf{A}^{\sigma_i} = \begin{pmatrix} 
\mathbf{A}^{\sigma_i,1} & 0  & ....& 0 \\
0 & \mathbf{A}^{\sigma_i,2} & ... & 0\\ 
 \vdots &  \vdots & \ddots & \vdots\\
0 & 0 &  ... & \mathbf{A}^{\sigma_i,k}\\ 
\end{pmatrix} , \label{eq:combin}\end{equation}
where the batch index runs from 1 to $k$. 
A subsequent sequence of SVDs would again lead to an MPS with reduced bond order and full rank.
For example, the results in Fig. \ref{fig:h8mps} were obtained by dividing the $50,000$ trajectories into 50 batches of size 1000. After the initial compression, the batches were combined in pairs according to (\ref{eq:combin}) and compressed again. This was repeated two more times in groups of five batches. In practice, if the bond dimension after a compression exceeds the nominal value of 20 we truncate it back to this value. 
Since the mapping to MPS must be carried out at each time where observables are calculated, it is more efficient to only carry it out at the times of interest. 

For all the MPO $W^{I}$ \cite{Zaletel2015} simulations carried out in the main text we use a maximum bond dimension of $50$, a minimum singular value cut-off  of $10^{-14}$, and a time-step of $dt = 0.001$.
 
\section{IV. Scaling} \label{sec:scaling}
\begin{figure}[h]
\subfloat[]{\includegraphics[width = 7cm]{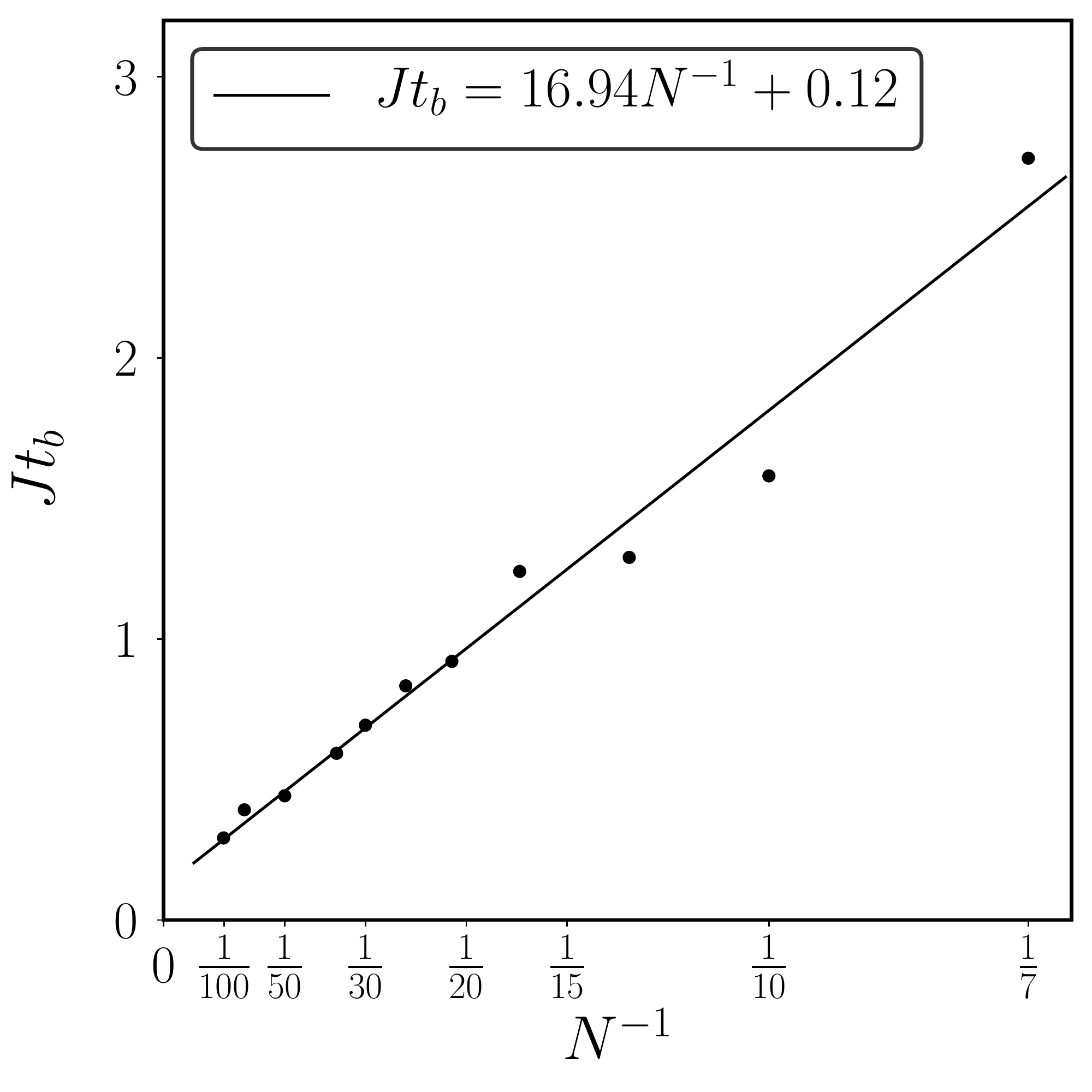}}
\hspace{0.5cm}
\subfloat[]{\includegraphics[width = 7.2cm]{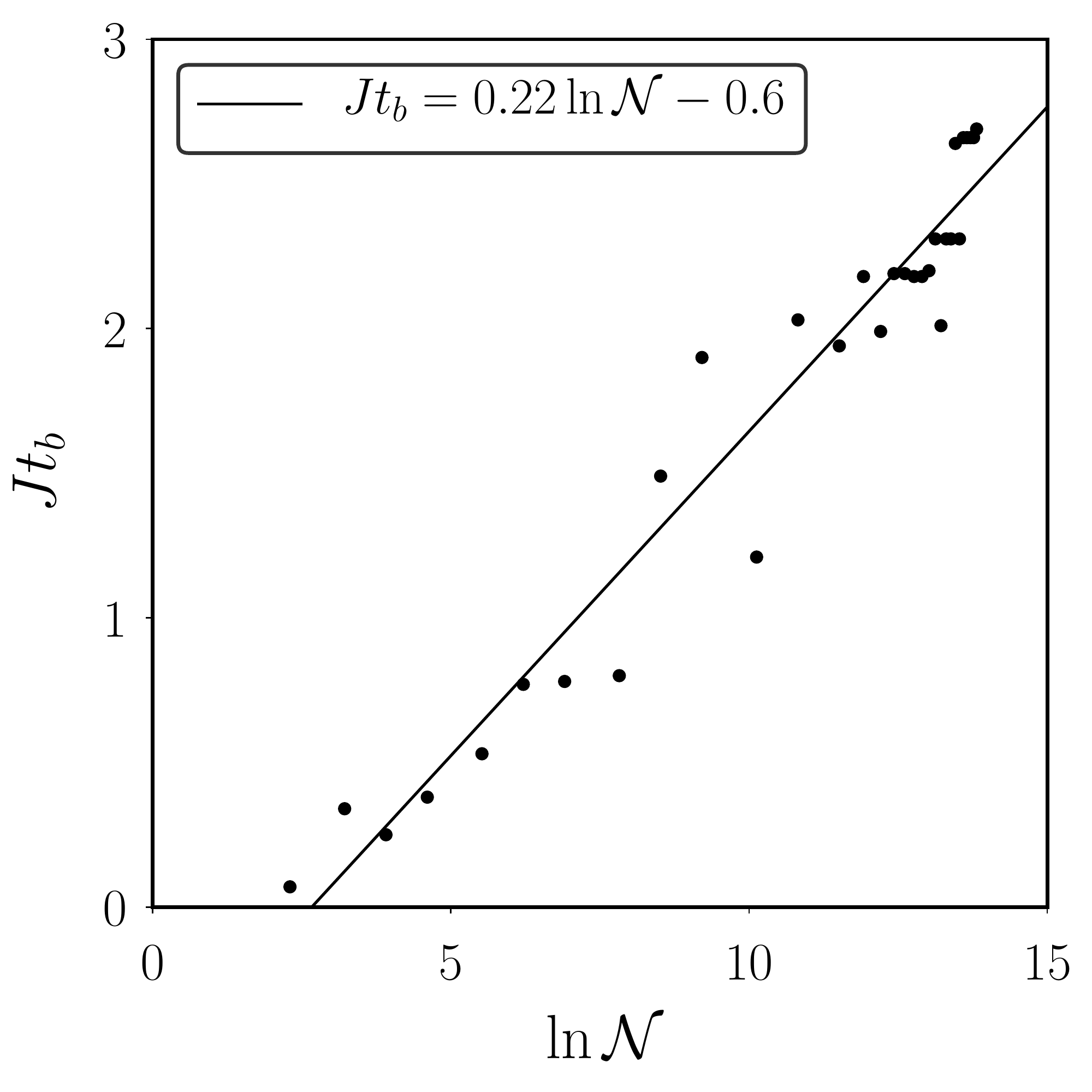}}
\caption{(a) Breakdown time, $t_b$, of the simulations following a quantum quench in the 1D quantum Ising model from the fully polarized initial state $\ket{\Downarrow}$ to $\Gamma = 8J$. The breakdown time is defined as the time at which the norm deviates by $10\%$ from unity. (a) Scaling of $t_b$  with inverse system size with $\mathcal{N} = 10^6$ held fixed. The data are well approximated by a linear fit (solid line), particularly for large system sizes. (b) Scaling of $t_b$ with the number of samples for a fixed system size with $N = 7$. The linear fit (solid line) suggests an exponential dependence of the number of samples on the $t_b$ according to $\mathcal{N} \propto e^{\alpha Jt_b}$, with $\alpha \approx 4.5$. }
\label{fig:scaling}
\end{figure}

In order to quantify the scaling properties of the real-time stochastic approach, we consider the time-scale over which the simulations are accurate as a function of the system size and the number of samples. Since the stochastic approach only produces perfectly normalized quantum states in the limit $\mathcal{N} \rightarrow \infty$, deviations of the norm from unity can provide an estimate of the simulation's convergence and the time-scale over which the method can be trusted. As illustrated in Fig. \ref{fig:typ}, rescaling by the norm can lead to good approximations for physical observables, even if the norm deviates significantly from unity. In view of this, we define the breakdown time, $t_b$, as the earliest time for which a $10\%$ error is observed in the norm. In Fig. \ref{fig:scaling}(a) we show $t_b$ as a function of the inverse system size $N^{-1}$, for quenches in the 1D quantum Ising model from the fully polarized initial state $\ket{\Downarrow}$ to $\Gamma = 8J$, for a fixed number of samples $\mathcal{N} = 10^6$. The data are well approximated by the linear relation
\begin{equation} 
Jt_b = 16.94 N^{-1} + 0.12 .
\end{equation} 
That is to say, for a fixed number of samples the breakdown time scales with the inverse of the system size. This is consistent with the results of \cite{DeNicola2019long}.
 In Fig. \ref{fig:scaling}(b) we also show $t_b$ as a function of the number of samples, for the same quench, but for a fixed system size, with $N = 7$ spins. The data are compatible with the relation $Jt_b = 0.22 \ln \mathcal{N} - 0.6$, i.e. the number of samples required to reach a given time $t_b$ therefore scales exponentially, $\mathcal{N} \propto e^{\alpha Jt_b}$, where $\alpha \approx 4.5$, in this example. This is consistent with the scaling of fluctuations analyzed in \cite{DeNicola2019long} using different diagnostics.

\end{document}